\documentclass{article}
\usepackage{amsmath,amssymb}

\newtheorem{theorem}{Theorem}[section]

\begin{document}

\title{Upper Bounds on the Probability of Error in terms of Mean Divergence Measures \thanks{%
Mathematics Subject Classifications: 94A17; 62B10.}}
\author{Inder Jeet Taneja \thanks{%
Departamento de Matem\'{a}tica, Universidade Federal de Santa Catarina,  88.040-900 Florian\'{o}polis, SC, Brazil}}
\maketitle

\begin{abstract}
In this paper we shall consider some famous means such as arithmetic, harmonic, geometric, root square mean, etc. Considering the difference of these means, we can establish \cite{tan2, tan3}. some inequalities among them. Interestingly, the difference of mean considered is convex functions. Applying some properties, upper bounds on the probability of error are established in this paper. It is also shown that the results obtained are sharper than obtained directly applying known inequalities.
\end{abstract}

\section{Introduction}

Taneja \cite{tan3, tan4} considered the following inequality among some well-known means:
\begin{equation}
\label{eq1}
H(a,b) \le G(a,b) \le N_1 (a,b)
 \le N_3 (a,b) \le N_2 (a,b) \le A(a,b) \le S(a,b),
\end{equation}

\noindent where
\begin{align}
A(a,b) & = \frac{a + b}{2},\notag\\
G(a,b) & = \sqrt {ab} ,\notag\\
H(a,b) & = \frac{2ab}{a + b},\notag\\
N_1 (a,b) & = \left( {\frac{\sqrt a + \sqrt b }{2}} \right)^2,\notag\\
N_2 (a,b) & = \left( {\frac{\sqrt a + \sqrt b }{2}} \right)\left( {\sqrt
{\frac{a + b}{2}} } \right),\notag\\
N_3 (a,b) &  = \frac{a + \sqrt {ab} + b}{3}\notag
\intertext{and}
S(a,b) & = \sqrt {\frac{a^2 + b^2}{2}}\notag
\end{align}

\noindent
for all $a,\;b \in (0,\infty )$

\smallskip
The means, $H(a,b)$, $G(a,b)$, $A(a,b)$ and $S(a,b)$ are known in the literature as \textit{harmonic}, \textit{geometric}, \textit{arithmetic} and \textit{root-square means} respectively. For simplicity, we can call the mean, $N_1 $ as \textit{square-root} mean. The $N_2 (a,b)$ can be seen in Taneja \cite{tan2}  and the mean $N_3 (a,b)$ can be seen in Zhang and Wu \cite{zhw}. Schur-geometric convexity of the  means appearing in (\ref{eq1}) can be seen in \cite{szd, sjz}

\smallskip
The above measures can be written in terms of \textit{arithmetic} and \textit{geometric means}. See below
\begin{align}
N_1 (a,b) & = \left( {A\left( {\sqrt a ,\sqrt b } \right)} \right)^2 =
\frac{A(a,b) + G(a,b)}{2},\notag\\
N_2 (a,b) & = A\left( {\sqrt a ,\sqrt b } \right) \cdot \sqrt {A(a,b)} ,\notag\\
N_3 (a,b) & = \frac{2\,A(a,b) + G(a,b)}{3}, \notag\\
H(a,b) & = \frac{\left( {G(a,b)} \right)^2}{A(a,b)}\notag
\intertext{and}
S(a,b) & = \sqrt {A\left( {a^2,b^2} \right)} .\notag
\end{align}

\section{Inequalities among difference of means}

Let us consider the following nonnegative \textbf{difference of means}:
\begin{equation}
\label{eq2}
M_{SA} (a,b) = S(a,b) - A(a,b),
\end{equation}
\begin{equation}
\label{eq3}
M_{SN_2 } (a,b) = S(a,b) - N_2 (a,b),
\end{equation}
\begin{equation}
\label{eq4}
M_{SN_3 } (a,b) = S(a,b) - N_3 (a,b),
\end{equation}
\begin{equation}
\label{eq5}
M_{SN_1 } (a,b) = S(a,b) - N_1 (a,b),
\end{equation}
\begin{equation}
\label{eq6}
M_{SG} (a,b) = S(a,b) - G(a,b),
\end{equation}
\begin{equation}
\label{eq7}
M_{SH} (a,b) = S(a,b) - H(a,b),
\end{equation}
\begin{equation}
\label{eq8}
M_{AN_2 } (a,b) = A(a,b) - N_2 (a,b),
\end{equation}
\begin{equation}
\label{eq9}
M_{AG} (a,b) = A(a,b) - G(a,b),
\end{equation}
\begin{equation}
\label{eq10}
M_{AH} (a,b) = A(a,b) - H(a,b),
\end{equation}
\begin{equation}
\label{eq11}
M_{N_2 N_1 } (a,b) = N_2 (a,b) - N_1 (a,b)
\end{equation}

\noindent and
\begin{equation}
\label{eq12}
M_{N_2 G} (a,b) = N_2 (a,b) - G(a,b).
\end{equation}

The convexity of the means (\ref{eq2})-(\ref{eq12}) can be seen in Taneja \cite{tan2, tan3, tan4}. The
Schur-geometric convexity is given in \cite{szd}. The mean difference $M_{N_2
N_3 } (a,b)$ is not considered here, since it is not convex.

\smallskip
Taneja  \cite{tan3} proved the following inequalities among the difference of means:
\begin{equation}
\label{eq13}
M_{SA} (a,b) \le \frac{1}{3}M_{SH} (a,b) \le \frac{1}{2}M_{AH} (a,b) \le
\frac{1}{2}M_{SG} (a,b) \le M_{AG} (a,b),
\end{equation}
\begin{equation}
\label{eq14}
\frac{1}{8}M_{AH} (a,b) \le M_{N_2 N_1 } (a,b) \le \frac{1}{3}M_{N_2 G}
(a,b) \le \frac{1}{4}M_{AG} (a,b) \le M_{AN_2 } (a,b),
\end{equation}
\begin{equation}
\label{eq15}
\frac{1}{4}M_{SA} (a,b) \le \frac{1}{5}M_{SN_2 } (a,b) \le M_{AN_2 } (a,b),
\end{equation}
\begin{equation}
\label{eq16}
\frac{1}{2}M_{SH} (a,b) \le M_{SN_1 } (a,b) \le \frac{3}{4}M_{SG} (a,b)
\end{equation}

\noindent and
\begin{equation}
\label{eq17}
M_{SA} (a,b) \le \frac{3}{4}M_{SN_3 } (a,b) \le \frac{2}{3}M_{SN_1 } (a,b).
\end{equation}

The aim of this paper is to obtain bounds on the probability of error in
terms of the means given in (\ref{eq2})-(\ref{eq12}).

\section{$f-$Divergence and Probability of Error}

Csisz\'{a}r \cite{csi} have given a measure for the divergence between two
probability density functions, say $p(x)$ and $q(x)$. This so called $f -
$divergence given by
\begin{equation}
\label{eq18}
C_f (p,q) = \int\limits_{\rm X} {f\left( {\frac{p(x)}{q(x)}} \right)q(x)dx}.
\end{equation}

The function $f(x)$, with $x \in (0,\infty )$ is a convex function which has
to satisfy the conditions
\begin{equation}
\label{eq19}
f(0) = \mathop {\lim }\limits_{u \downarrow 0} f(u);
\quad
0f\left( {\frac{0}{0}} \right) = 0 \quad ;
\quad
0f\left( {\frac{a}{0}} \right) = \mathop {\lim }\limits_{ \in \downarrow
\infty } \in f\left( {\frac{a}{ \in }} \right) = a\mathop {\lim }\limits_{x
\to \infty } \frac{f(u)}{u}.
\end{equation}

It can be easily checked that $C_f (p,q) \ge f(1)$ and that $C_f (p,q) =
f(1)$ only when $p(x) = q(x)$ a.e. Thus, $C_f (p,q) - f(1)$ is a distance or
divergence measure in the sense that $C_f (p,q) - f(1) \ge 0$. However, it
is not symmetric in $p$ and $q$ and in general does not satisfy triangle
inequality.

\smallskip
Boekee and Van der Lubbe \cite{bov} have introduced the average $f -$divergence between two hypothesis $C_1 $ and $C_2 $ in terms of their ``\textit{a posteriori}'' probabilities. This average $f - $divergence is defined as
\begin{equation}
\label{eq20}
C_f (C_1 ,C_2 ) = \int\limits_{\rm X} {f\left( {\frac{P(C_1 \vert x)}{P(C_2
\vert x)}} \right)P(C_1 \vert x)p(x)dx}
 = E_X \left\{ {f\left( {\frac{P(C_1 \vert x)}{P(C_2 \vert x)}} \right)P(C_1
\vert x)} \right\}
\end{equation}

If we introduce the function
\[
f^ * (u) = u\,f\left( {\frac{1 - u}{u}} \right)
\]

\noindent
and set $u = u(x) = P(C_2 \vert x)$, it is easy to see from $P(C_1 \vert x)
= 1 - P(C_2 \vert x)$ that
\begin{equation}
\label{eq21}
\overline C _f (C_1 ,C_2 ) = \int\limits_{\rm X} {f^ * \left( {\frac{P(C_1
\vert x)}{P(C_2 \vert x)}} \right)P(C_1 \vert x)p(x)dx} = E_X \left\{ {f^ *
\left( {\frac{P(C_1 \vert x)}{P(C_2 \vert x)}} \right)P(C_1 \vert x)}
\right\}.
\end{equation}

From Vajda \cite{vaj} it follows that $f^ * (u)$ is convex on $[0,1]$ and is
strictly convex iff $f(u)$ is strictly convex.

\subsection{A Class of Upper Bounds}

In \cite{bov} it has been shown that the Bayesian probability of error can be
upper bounds in terms of the average $f - $\textit{divergence} $\overline C _f (C_1 ,C_2 )$.
This upper bound is given by
\begin{equation}
\label{eq22}
P_e \le \frac{f_0 P(C_2 ) + f_\infty P(C_1 ) - \overline C _f (C_1 ,C_2
)}{f_2 - f_1 },
\end{equation}

\noindent
where $f_2 $ should be finite with
\begin{equation}
\label{eq23}
f_0 = \mathop {\lim }\limits_{u \downarrow \infty } f(u);
\quad
f_1 = f(1);
\quad
f_2 = f_0 + f_\infty ;
\quad
f_\infty = \mathop {\lim }\limits_{u \to \infty } \frac{f(u)}{u}.
\end{equation}

The above bound is valid only for every \textit{convex function} $f(u)$ which satisfies the
conditions given in (\ref{eq19}). However, if $f^ * (u) = u\,f\left( {\frac{1 -
u}{u}} \right)$ is symmetric with respect to $u = \frac{1}{2}$ i.e., $f^ *
(u) = f^ * (1 - u)$, this bound can be written in a simpler form given in
the following theorem.

\begin{theorem} The probability of error is upper bounds by $\overline
C _f (C_1 ,C_2 )$, where $f^ * (u) = f^ * (1 - u)$ is symmetric with respect
to $u = \frac{1}{2}$, as follows:
\begin{equation}
\label{eq24}
P_e \le \frac{1}{2f_\infty - f_1 }\left[ {f_\infty - \overline C _f (C_1
,C_2 )} \right],
\end{equation}

\noindent
provided $f_\infty $ is finite.
\end{theorem}

If $f_1 = f(1) = 0$ and $f_\infty $ is finite, then the above bound can be
written as
\begin{equation}
\label{eq25}
P_e \le \frac{1}{2}\left[ {1 - \frac{1}{f_\infty }\overline C _f (C_1 ,C_2
)} \right].
\end{equation}

In this paper we shall apply the upper bound (\ref{eq25}) for different divergences
based on difference of means given by (\ref{eq2})-(\ref{eq12}) .

\section{Bounds on the Probability of Error in terms of Difference Mean Divergences}

In this section we shall give bounds on the probability of error in terms of
the mean differences given by (\ref{eq2})-(\ref{eq12}) based the Theorem 3.1.

\smallskip
\textbf{Result 4.1. } Let us consider the measure
\[
\overline M _{SA} (C_1 ,C_2 ) = E_X \left\{ {f_{SA}^ * \left( {\frac{P(C_1
\vert x)}{P(C_2 \vert x)}} \right)P(C_1 \vert x)} \right\},
\]

\noindent where
\[
f_{SA}^ * (x) = x\,f_{SA} \left( {\frac{1 - x}{x}} \right)
\]

\noindent with
\[
f_{SA} (x) = \sqrt {\frac{x^2 + 1}{2}} - \frac{x + 1}{2},\;\forall x \in
(0,\infty ).
\]

The convexity of the function $f_{SA} (x)$ can be seen in \cite{tan2, tan3}. In this
case we have
\begin{equation}
\label{eq26}
f_{SA}^ * (x) = \frac{1}{2}\left[ {\sqrt {2\left( {x^2 + (1 - x)^2} \right)}
- 1} \right] = f_{SA}^ * (1 - x),
\end{equation}
\begin{equation}
\label{eq27}
f_{SA_\infty } = \mathop {\lim }\limits_{x \to \infty } \frac{f_{SA} (x)}{x}
= \frac{1}{2}\left( {\sqrt 2 - 1} \right)
\end{equation}

\noindent and
\begin{equation}
\label{eq28}
f_{SA} (1) = 0.
\end{equation}

Expression (\ref{eq25}) together with (\ref{eq26})-(\ref{eq28}) give the following upper bound
on the probability of error
\begin{equation}
\label{eq29}
P_e \le \frac{1}{2}\left[ {1 - \left( {\frac{2}{\sqrt 2 - 1}}
\right)\overline M _{SA} (C_1 ,C_2 )} \right].
\end{equation}

\textbf{Result 4.2. }Let us consider the measure
\[
\overline M _{SN_2 } (C_1 ,C_2 ) = E_X \left\{ {f_{SN_2 }^\ast \left(
{\frac{P(C_1 \vert x)}{P(C_2 \vert x)}} \right)P(C_1 \vert x)} \right\},
\]

\noindent where
\[
f_{SN_2 }^ * (x) = x\,f_{SN_2 } \left( {\frac{1 - x}{x}} \right)
\]

\noindent with
\[
f_{SN_2 } (x) = \sqrt {\frac{x^2 + 1}{2}} - \left( {\frac{\sqrt x + 1}{2}}
\right)\sqrt {\frac{x + 1}{2}} ,\;\forall x \in (0,\infty ).
\]

The convexity of the function $f_{SN_2 } (x)$ can be seen in\cite{tan2, tan3}. In
this case we have
\begin{equation}
\label{eq30}
f_{SN_2 }^ * (x) = \frac{\sqrt 2 }{4}\left( {2\sqrt {x^2 + (1 - x)^2} -
\sqrt x - \sqrt {1 - x} } \right) = f_{SN_2 }^ * (1 - x),
\end{equation}
\begin{equation}
\label{eq31}
f_{\left( {SN_2 } \right)_\infty } = \mathop {\lim }\limits_{x \to \infty }
\frac{f_{SN_2} (x)}{x} = \frac{\sqrt 2 }{4}
\end{equation}

\noindent and
\begin{equation}
\label{eq32}
f_{SN_2 } (1) = 0.
\end{equation}

Expression (\ref{eq25}) together with (\ref{eq30})-(\ref{eq32}) give the following upper bound
on the probability of error
\begin{equation}
\label{eq33}
P_e \le \frac{1}{2}\left[ {1 - \frac{4}{\sqrt 2 }\overline M _{SN_2 } (C_1
,C_2 )} \right].
\end{equation}

\textbf{Result 4.3. }Let us consider the measure
\[
\overline M _{SN_3 } (C_1 ,C_2 ) = E_X \left\{ {f_{SN_3 }^\ast \left(
{\frac{P(C_1 \vert x)}{P(C_2 \vert x)}} \right)P(C_1 \vert x)} \right\},
\]

\noindent where
\[
f_{SN_3 }^ * (x) = x\,f_{SN_3 } \left( {\frac{1 - x}{x}} \right)
\]

\noindent with
\[
f_{SN_3 } (x) = \sqrt {\frac{x^2 + 1}{2}} - \frac{x + \sqrt x +
1}{3},\;\forall x \in (0,\infty ).
\]

The convexity of the function $f_{SN_3 } (x)$ can be seen in \cite{tan2, tan3}. In
this case we have
\begin{equation}
\label{eq34}
f_{SN_3 }^ * (x) = \frac{\sqrt 2 }{2}\sqrt {x^2 + (1 - x)^2} -
\frac{1}{3}\left( {1 + \sqrt {x(1 - x)} } \right) = f_{SN_3 }^ * (1 - x),
\end{equation}
\begin{equation}
\label{eq35}
f_{\left( {SN_3 } \right)_\infty } = \mathop {\lim }\limits_{x \to \infty }
\frac{f_{SA} (x)}{x} = \frac{\sqrt 2 }{2} - \frac{1}{3} = \frac{3\sqrt 2 -
2}{6}
\end{equation}

\noindent and
\begin{equation}
\label{eq36}
f_{SN_3 } (1) = 0.
\end{equation}

Expression (\ref{eq25}) together with (\ref{eq34})-(\ref{eq36}) give the following upper bound
on the probability of error
\begin{equation}
\label{eq37}
P_e \le \frac{1}{2}\left[ {1 - \frac{6}{3\sqrt 2 - 2}\overline M _{SN_3 }
(C_1 ,C_2 )} \right].
\end{equation}

\textbf{Result 4.4. }Let us consider the measure
\[
\overline M _{SN_1 } (C_1 ,C_2 ) = E_X \left\{ {f_{SN_1 }^\ast \left(
{\frac{P(C_1 \vert x)}{P(C_2 \vert x)}} \right)P(C_1 \vert x)} \right\},
\]

\noindent where
\[
f_{SN_1 }^ * (x) = x\,f_{SN_1 } \left( {\frac{1 - x}{x}} \right),
\]

\noindent with
\[
f_{SN_1 } (x) = \sqrt {\frac{x^2 + 1}{2}} - \left( {\frac{\sqrt x + 1}{2}}
\right)^2,\;\forall x \in (0,\infty ).
\]

The convexity of the function $f_{SN_1 } (x)$ can be seen in \cite{tan2, tan3}. In
this case we have
\begin{equation}
\label{eq38}
f_{SN_1 }^ * (x) = \frac{\sqrt 2 }{2}\sqrt {x^2 + (1 - x)^2} -
\frac{1}{4}\left( {1 + 2\sqrt {x(1 - x)} } \right) = f_{SN_1 }^ * (1 - x),
\end{equation}
\begin{equation}
\label{eq39}
f_{\left( {SN_1 } \right)_\infty } = \mathop {\lim }\limits_{x \to \infty }
\frac{f_{SN_1 } (x)}{x} = \frac{\sqrt 2 }{2} - \frac{1}{4} = \frac{2\sqrt 2
- 1}{4}
\end{equation}

\noindent and
\begin{equation}
\label{eq40}
f_{SN_1 } (1) = 0.
\end{equation}

Expression (\ref{eq25}) together with (\ref{eq38})-(\ref{eq40}) give the following upper bound
on the probability of error
\begin{equation}
\label{eq41}
P_e \le \frac{1}{2}\left[ {1 - \frac{4}{2\sqrt 2 - 1}\overline M _{SN_1 }
(C_1 ,C_2 )} \right].
\end{equation}

\textbf{Result 4.5. }Let us consider the measure
\[
\overline M _{SG} (C_1 ,C_2 ) = E_X \left\{ {f_{SG}^\ast \left( {\frac{P(C_1
\vert x)}{P(C_2 \vert x)}} \right)P(C_1 \vert x)} \right\},
\]

\noindent where
\[
f_{SG}^ * (x) = x\,f_{SG} \left( {\frac{1 - x}{x}} \right),
\]

\noindent with
\[
f_{SG} (x) = \sqrt {\frac{x^2 + 1}{2}} - \sqrt x ,\;\forall x \in (0,\infty
).
\]

The convexity of the function $f_{SG} (x)$ can be seen in \cite{tan2, tan3}. In this
case we have
\begin{equation}
\label{eq42}
f_{SG}^ * (x) = \frac{\sqrt 2 }{2}\sqrt {x^2 + (1 - x)^2} - \sqrt {x(1 - x)}
= f_{SG}^ * (1 - x),
\end{equation}
\begin{equation}
\label{eq43}
f_{\left( {SG} \right)_\infty } = \mathop {\lim }\limits_{x \to \infty }
\frac{f_{SG} (x)}{x} = \frac{\sqrt 2 }{2}
\end{equation}

\noindent and
\begin{equation}
\label{eq44}
f_{SG} (1) = 0.
\end{equation}

Expression (\ref{eq25}) together with (\ref{eq41})-(\ref{eq43}) give the following upper bound
on the probability of error
\begin{equation}
\label{eq45}
P_e \le \frac{1}{2}\left[ {1 - \frac{2}{\sqrt 2 }\overline M _{SG} (C_1 ,C_2
)} \right].
\end{equation}

\textbf{Result 4.6. }Let us consider the measure
\[
\overline M _{SH} (C_1 ,C_2 ) = E_X \left\{ {f_{SH}^\ast \left( {\frac{P(C_1
\vert x)}{P(C_2 \vert x)}} \right)P(C_1 \vert x)} \right\},
\]

\noindent where
\[
f_{SH}^ * (x) = x\,f_{SH} \left( {\frac{1 - x}{x}} \right)
\]

\noindent with
\[
f_{SH} (x) = \sqrt {\frac{x^2 + 1}{2}} - \frac{2x}{x + 1},\;\forall x \in
(0,\infty ).
\]

The convexity of the function $f_{SH} (x)$ can be seen in \cite{tan2, tan3}. In this
case we have
\begin{equation}
\label{eq46}
f_{SH}^ * (x) = \frac{\sqrt 2 }{2}\sqrt {x^2 + (1 - x)^2} - 2x(1 - x) =
f_{SH}^ * (1 - x),
\end{equation}
\begin{equation}
\label{eq47}
f_{\left( {SH} \right)_\infty } = \mathop {\lim }\limits_{x \to \infty }
\frac{f_{SH} (x)}{x} = \frac{\sqrt 2 }{2}
\end{equation}

\noindent and
\begin{equation}
\label{eq48}
f_{SH} (1) = 0.
\end{equation}

Expression (\ref{eq25}) together with (\ref{eq46})-(\ref{eq48}) give the following upper bound
on the probability of error
\begin{equation}
\label{eq49}
P_e \le \frac{1}{2}\left[ {1 - \frac{2}{\sqrt 2 }\overline M _{SH} (C_1 ,C_2
)} \right].
\end{equation}

\textbf{Result 4.7. }Let us consider the measure
\[
\overline M _{AN_2 } (C_1 ,C_2 ) = E_X \left\{ {f_{AN_2 }^\ast \left(
{\frac{P(C_1 \vert x)}{P(C_2 \vert x)}} \right)P(C_1 \vert x)} \right\},
\]

\noindent where
\[
f_{AN_2 }^ * (x) = x\,f_{AN_2 } \left( {\frac{1 - x}{x}} \right)
\]

\noindent with
\[
f_{AN_2 } (x) = \frac{x + 1}{2} - \left( {\frac{\sqrt x + 1}{2}}
\right)\sqrt {\frac{x + 1}{2}} ,\;\forall x \in (0,\infty ).
\]

The convexity of the function $f_{AN_2 } (x)$ can be seen in \cite{tan2, tan3}. In
this case we have
\begin{equation}
\label{eq50}
f_{AN_2 }^ * (x) = \frac{1}{2} - \frac{\sqrt 2 }{4}\left( {\sqrt x + \sqrt
{1 - x} } \right) = f_{AN_2 }^ * (1 - x),
\end{equation}
\begin{equation}
\label{eq51}
f_{\left( {AN_2 } \right)_\infty } = \mathop {\lim }\limits_{x \to \infty }
\frac{f_{AN_2 } (x)}{x} = \frac{1}{2} - \frac{\sqrt 2 }{4} = \frac{2 - \sqrt
2 }{4}
\end{equation}

\noindent and
\begin{equation}
\label{eq52}
f_{AN_2 } (1) = 0.
\end{equation}

Expression (\ref{eq25}) together with (\ref{eq50})-(\ref{eq52}) give the following upper bound
on the probability of error
\begin{equation}
\label{eq53}
P_e \le \frac{1}{2}\left[ {1 - \left( {\frac{4}{2 - \sqrt 2 }}
\right)\overline M _{AN_2 } (C_1 ,C_2 )} \right].
\end{equation}

\textbf{Result 4.8. }Let us consider the measure
\[
\overline M _{AG} (C_1 ,C_2 ) = E_X \left\{ {f_{AG}^\ast \left( {\frac{P(C_1
\vert x)}{P(C_2 \vert x)}} \right)P(C_1 \vert x)} \right\},
\]

\noindent where
\[
f_{AG}^ * (x) = x\,f_{AG} \left( {\frac{1 - x}{x}} \right),
\]

\noindent with
\[
f_{AG} (x) = \frac{x + 1}{2} - \sqrt x ,\;\forall x \in (0,\infty ).
\]

The convexity of the function $f_{AG} (x)$ can be seen in \cite{tan2, tan3}. In this
case we have
\begin{equation}
\label{eq54}
f_{AG}^ * (x) = \frac{1}{2} - \sqrt {x(1 - x)} = f_{AG}^ * (1 - x),
\end{equation}
\begin{equation}
\label{eq55}
f_{\left( {AG} \right)_\infty } = \mathop {\lim }\limits_{x \to \infty }
\frac{f_{AG} (x)}{x} = \frac{1}{2}
\end{equation}

\noindent and
\begin{equation}
\label{eq56}
f_{AG} (1) = 0.
\end{equation}

Expression (\ref{eq25}) together with (\ref{eq52})-(\ref{eq54}) give the following upper bound
on the probability of error
\begin{equation}
\label{eq57}
P_e \le \frac{1}{2}\left[ {1 - 2\;\overline M _{AG} (C_1 ,C_2 )} \right].
\end{equation}

\textbf{Result 4.9. }Let us consider the measure
\[
\overline M _{AH} (C_1 ,C_2 ) = E_X \left\{ {f_{AH}^\ast \left( {\frac{P(C_1
\vert x)}{P(C_2 \vert x)}} \right)P(C_1 \vert x)} \right\},
\]

\noindent where
\[
f_{AH}^ * (x) = x\,f_{AH} \left( {\frac{1 - x}{x}} \right),
\]

\noindent with
\[
f_{AH} (x) = \frac{x + 1}{2} - \frac{2x}{x + 1},\;\forall x \in (0,\infty
).
\]

The convexity of the function $f_{AH} (x)$ can be seen in \cite{tan2, tan3}. In this
case we have
\begin{equation}
\label{eq58}
f_{AH}^ * (x) = \frac{1}{2}(2x - 1)^2 = f_{AH}^ * (1 - x),
\end{equation}
\begin{equation}
\label{eq59}
f_{\left( {AH} \right)_\infty } = \mathop {\lim }\limits_{x \to \infty }
\frac{f_{AH} (x)}{x} = \frac{1}{2}
\end{equation}

\noindent and
\begin{equation}
\label{eq60}
f_{AH} (1) = 0.
\end{equation}

Expression (\ref{eq25}) together with (\ref{eq58})-(\ref{eq60}) give the following upper bound
on the probability of error
\begin{equation}
\label{eq61}
P_e \le \frac{1}{2}\left[ {1 - 2\;\overline M _{AH} (C_1 ,C_2 )} \right].
\end{equation}

\textbf{Result 4.10. }Let us consider the measure
\[
\overline M _{N_2 N_1 } (C_1 ,C_2 ) = E_X \left\{ {f_{N_2 N_1 }^\ast \left(
{\frac{P(C_1 \vert x)}{P(C_2 \vert x)}} \right)P(C_1 \vert x)} \right\},
\]

Where
\[
f_{N_2 N_1 }^ * (x) = x\,f_{N_2 N_1 } \left( {\frac{1 - x}{x}} \right),
\]

\noindent with
\[
f_{N_2 N_1 } (x) = \left( {\frac{\sqrt x + 1}{2}} \right)\sqrt {\frac{x +
1}{2}} - \left( {\frac{\sqrt x + 1}{2}} \right)^2,\;\forall x \in (0,\infty
).
\]

The convexity of the function $f_{N_2 N_1 } (x)$ can be seen in \cite{tan2, tan3}.
In this case we have
\begin{equation}
\label{eq62}
f_{N_2 N_1 }^ * (x) = \frac{\sqrt 2 }{4}\left( {\sqrt x + \sqrt {1 - x} }
\right) - \frac{1}{4}\left( {1 + 2\sqrt {x(1 - x)} } \right) = f_{N_2 N_1 }^
* (1 - x),
\end{equation}
\begin{equation}
\label{eq63}
f_{\left( {N_2 N_1 } \right)_\infty } = \mathop {\lim }\limits_{x \to \infty
} \frac{f_{N_2 N_1 } (x)}{x} = \frac{\sqrt 2 }{2} - \frac{1}{4} =
\frac{2\sqrt 2 - 1}{4}
\end{equation}

\noindent and
\begin{equation}
\label{eq64}
f_{N_2 N_1 } (1) = 0.
\end{equation}

Expression (\ref{eq25}) together with (\ref{eq61})-(\ref{eq64}) give the following upper bound
on the probability of error
\begin{equation}
\label{eq65}
P_e \le \frac{1}{2}\left[ {1 - \frac{4}{2\sqrt 2 - 1}\overline M _{N_2 N_1 }
(C_1 ,C_2 )} \right].
\end{equation}

\textbf{Result 4.11. }Let us consider the measure
\[
\overline M _{N_2 G} (C_1 ,C_2 ) = E_X \left\{ {f_{N_2 G}^\ast \left(
{\frac{P(C_1 \vert x)}{P(C_2 \vert x)}} \right)P(C_1 \vert x)} \right\},
\]

\noindent where
\[
f_{N_2 G}^ * (x) = x\,f_{N_2 G} \left( {\frac{1 - x}{x}} \right),
\]

\noindent with
\[
f_{N_2 G} (x) = \left( {\frac{\sqrt x + 1}{2}} \right)\sqrt {\frac{x +
1}{2}} - \sqrt x ,\;\forall x \in (0,\infty ).
\]

The convexity of the function $f_{N_2 G} (x)$ can be seen in \cite{tan2, tan3}. In
this case we have
\begin{equation}
\label{eq66}
f_{N_2 G}^ * (x) = \frac{\sqrt 2 }{4}\left( {\sqrt x + \sqrt {1 - x} }
\right) - \sqrt {x(1 - x)} = f_{N_2 G}^ * (1 - x),
\end{equation}
\begin{equation}
\label{eq67}
f_{\left( {N_2 G} \right)_\infty } = \mathop {\lim }\limits_{x \to \infty }
\frac{f_{N_2 G} (x)}{x} = \frac{\sqrt 2 }{4}
\end{equation}

\noindent and
\begin{equation}
\label{eq68}
f_{N_2 N_1 } (1) = 0.
\end{equation}

Expression (\ref{eq25}) together with (\ref{eq66})-(\ref{eq68}) give the following upper bound
on the probability of error
\begin{equation}
\label{eq69}
P_e \le \frac{1}{2}\left[ {1 - \frac{4}{\sqrt 2 }\overline M _{N_2 N_1 }
(C_1 ,C_2 )} \right].
\end{equation}

\subsection{Final Remarks}
\noindent
(i) According to inequalities (\ref{eq13}) and the result (\ref{eq57}), we have
\begin{align}
\label{eq70}
 P_e \le & \frac{1}{2}\left[ {1 - 2 \overline M _{AG }
(C_1 ,C_2 )} \right] \le \frac{1}{2}\left[
{1 - \overline M _{SG }
(C_1 ,C_2 )} \right] \le  \frac{1}{2}\left[ {1 - \overline M _{AH }
(C_1 ,C_2 )} \right] \le \notag\\
 & \le  \frac{1}{2}\left[ {1 - \frac{2}{3} \overline M _{SH }
(C_1 ,C_2 )} \right] \le
\frac{1}{2}\left[ {1 - 2 \overline M _{SA }
(C_1 ,C_2 )} \right]. 
 \end{align}

From (\ref{eq70}) and (\ref{eq45}), we have
\begin{equation}
\label{eq71}
P_e \le \frac{1}{2}\left[ {1 - \frac{2}{\sqrt 2 } \overline M _{SG }
(C_1 ,C_2 )} \right] \le
\frac{1}{2}\left[ {1 - \overline M _{SG }
(C_1 ,C_2 )} \right].
\end{equation}

Again from (\ref{eq70}) and (\ref{eq61}), we have
\begin{equation}
\label{eq72}
P_e \le \frac{1}{2}\left[ {1 - 2 \overline M _{AH }
(C_1 ,C_2 )} \right] \le \frac{1}{2}\left[
{1 - \overline M _{AH }
(C_1 ,C_2 )} \right].
\end{equation}

From the expressions (\ref{eq71}) and (\ref{eq72}) we observe that the results obtained
here individually are sharper than that we get from the inequalities given
in (\ref{eq13}).

\bigskip
\noindent
(ii) According to inequalities (\ref{eq16}) and the result (\ref{eq45}), we have
\begin{align}
\label{eq73}
P_e \le & \frac{1}{2}\left[ {1 - \frac{2}{\sqrt 2 } \overline M _{SG }
(C_1 ,C_2 )} \right] \le \notag\\
& \le \frac{1}{2}\left[ {1 - \frac{8}{3\sqrt 2 }\overline M _{SN_1 }
(C_1 ,C_2 )} \right] \le \frac{1}{2}\left[ {1 - \frac{4}{3\sqrt 2 }\overline M _{SH }
(C_1 ,C_2 )} \right].
\end{align}

From (\ref{eq73}) and (\ref{eq41}), we have
\begin{equation}
\label{eq74}
P_e \le \frac{1}{2}\left[ {1 - \frac{4}{2\sqrt 2 - 1}\overline M _{SN_1 }
(C_1 ,C_2 )} \right] \le \frac{1}{2}\left[ {1 - \frac{8}{3\sqrt 2 }\overline M _{SN_1 }
(C_1 ,C_2 )} \right].
\end{equation}

Again from (\ref{eq73}) and (\ref{eq49}) we have
\begin{equation}
\label{eq75}
P_e \le \frac{1}{2}\left[ {1 - \frac{2}{\sqrt 2 }\overline M _{SH }
(C_1 ,C_2 )} \right] \le \frac{1}{2}\left[ {1 - \frac{4}{3\sqrt 2 }\overline M _{SH }
(C_1 ,C_2 )} \right].
\end{equation}

From the expressions (\ref{eq74}) and (\ref{eq75}), we observe that the results obtained
here individually are sharper than that we get from the inequalities given
in (\ref{eq16}).

\smallskip
Similarly, we can compare other results proving that the results obtained
individually are sharper than applying directly the inequalities given in
(\ref{eq13})-(\ref{eq17}).
More studies on probability of error having different entropy-type and generalized divergence measures can be seen in Taneja \cite{tan5, tan1}

\noindent
\textit{e-mail: ijtaneja@gmail.com\\
web-site: http://www.mtm.ufsc.br/$\sim$taneja}


\begin{thebibliography}{99}
\setlength{\itemsep}{5pt}

\bibitem{bov} D.E. BOEKEE and J.C. VAN DER LUBBE, Some Aspects of Error Bounds in Feature Selection, \textit{Pattern Recognition}, \textbf{11}(1979), 353-360.

\bibitem{csi} I. CSISZ\'{A}R, Information Type Measures of Differences of Probability Distribution and Indirect Observations, \textit{Studia Math. Hungarica}, \textbf{2}(1967), 299-318.

   \bibitem{tan5} I.J. TANEJA,  On Generalized Information Measures and Their Applications, Chapter in: \textit{Advances in Electronics and Electron Physics}, Ed. P.W. Hawkes, Academic Press, \textbf{76}(1989), 327-413.

\bibitem{tan1} I.J. TANEJA, New Developments in Generalized Information Measures, Chapter in: \textit{Advances in Imaging and Electron Physics}, Ed. P.W. Hawkes, \textbf{91}(1995), 37-136.

 \bibitem{tan2} I.J. TANEJA, On Symmetric and Nonsymmeric Divergence Measures and Their Generalizations, Chapter in: Advances in Imaging and Electron Physics, \textbf{138}(2005),      177-250.

  \bibitem{tan3} 	I.J. TANEJA, Refinement of Inequalities among Means,\textit{ Journal of Combinatorics, Information and Systems Sciences},  (special issue in honor of Prof. B.D. Sharma) Edited by Prof. Sat Gupta, Vol. 31(2006), 357-378.

\bibitem{tan4}  I.J. TANEJA, On Mean Divergence Measures,  \textit{Advances in Inequalities from Probability Theory and Statistics} - Edited by N.S. Barnett and S.S. Dragomir, Nova Science Publishers, 2008, 195-215.

\bibitem{vaj} VAJDA, I. On the $f-$divergence and singularity of probability measures, \textit{Periodica Math. Hunger}, \textbf{2}(1972), 223-234.

\bibitem{zhw} ZHI-HUA ZHANG and YU-DONG WU, The New Bounds of the Logarithmic Mean, \textit{RGMIA Research Report Collection}, \textit{http://rgmia.vu.edu.au}, \textbf{7}(2)(2004), Art. 7.

\bibitem{szd} H. N. SHI, J. ZHANG and DA-MAO LI, Schur-Geometric Convexity for Difference of Means, \textit{Applied Mathematics E-Notes}, \textbf{10}(2010), 275-284

\bibitem{sjz} HUAN NAN SHI and JIAN ZHANG, Schur-Geometric Concavity for Difference of Means,
available on line - \textit{http://rgmia.org/papers/v12n4/huannan2.pdf}, 2010, 1-12

\end{thebibliography}
\end{document}